\documentclass[aps,prb,showpacs,twocolumn,floats]{revtex4-1}
\usepackage{amssymb}
\usepackage{epsfig}
\usepackage{dcolumn}
\usepackage{bm}
\usepackage{amsmath}

\begin{document}

\title{Renormalization of the tunnel splitting in a rotating nanomagnet}
\bibliographystyle{prsty}
\author{Michael F. O'Keeffe and Eugene M. Chudnovsky}

\affiliation{Physics Department, Lehman College, City University of New York,
    250 Bedford Park Boulevard West, Bronx, New York, 10468-1589, USA}

\date{\today}

\begin{abstract}
We study spin tunneling in a magnetic nanoparticle with biaxial
anisotropy that is free to rotate about its anisotropy axis. Exact
instanton of the coupled equations of motion is found that
connects degenerate classical energy minima. We show that
mechanical freedom of the particle renormalizes magnetic
anisotropy and increases the tunnel splitting.
\end{abstract}

\pacs{75.45.+j, 75.75.Jn, 75.50.Xx}

\maketitle

Macroscopic dynamics of a fixed-length magnetic moment, ${\bf M}$,
of a single-domain ferromagnetic particle is described by the
Landau-Lifshitz equation \cite{LL-old,Lectures}. When dissipation
(which is usually weak) is neglected this equation reads
\begin{equation}
\frac{\partial \mathbf M}{\partial t} = \gamma \mathbf M \times
\mathbf B_{eff}, \;\;
    \mathbf B_{eff} = - \frac{\delta \mathcal E}{\delta \mathbf
    M}\,,
\label{eq:LL}
\end{equation}
where $\mathcal E$ is the classical magnetic energy of the
particle that depends on the orientation of ${\bf M}$. It was
shown long ago \cite{EC-JETP,EC-Gunther} that Eq.\ (\ref{eq:LL}),
besides the real-time solutions, also possesses imaginary-time
solutions - instantons - that describe macroscopic quantum
tunneling of ${\bf M}$ between classically degenerate energy
minima (see also books: Refs. \onlinecite{Lectures} and
\onlinecite{MQT}). In early experiments on spin tunneling
\cite{MQT} single-domain magnetic particles were always frozen in
a solid matrix so that their physical position and orientation
were fixed and only rotation of the magnetic moment was allowed.
Later, beams of small magnetic clusters were investigated
\cite{Cox,Heer,Bucher,Douglass,Billas,Xu} and more recently free
magnetic nanoparticles confined within solid nanocavities have
been studied \cite{Tejada-2010}. Experimentalists have also worked
with molecular nanomagnets deposited on surfaces
\cite{Zobbi,Martinez,Barraza,Pennino} or carbon nanotubes
\cite{Wern-CNT}, as well as with single magnetic molecules bridged
between metallic electrodes
\cite{Heersche,Jo,Henderson,Voss,Park}. In such experiments the
particles retain some mechanical freedom. This inspired recent
theoretical work on quantum mechanics of rotating magnets
\cite{CG-2010,Jaafar,LCT,Kovalev}. In this paper we obtain exact
magnetic instanton for a single-domain particle that is free to
rotate about its anisotropy axis.

\begin{figure}
\begin{center}
\includegraphics[width=90mm,angle=0]{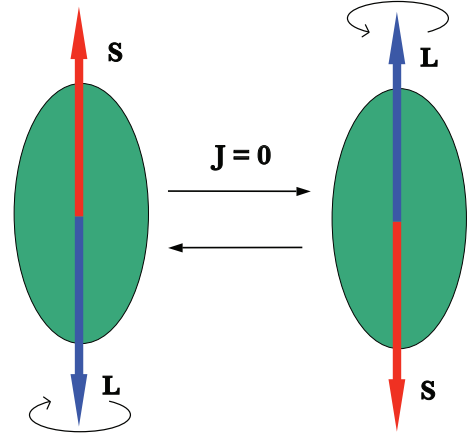}
\end{center}
\caption{Color online: Spin and angular momentum tunneling in a
rotating magnetic nanoparticle. } \label{fig:J0intro}
\end{figure}
General analytical solution for the rotational quantum levels of a
rigid body does not exist \cite{Edmonds}. Spin degree of freedom
makes this problem even less tractable. However, as was recently
demonstrated in Ref. \onlinecite{CG-2010}, the exact eigenstates
and exact energy levels can be obtained analytically for a
nanomagnet that, due to a large magnetic anisotropy, can be
described as a two-state spin system and is free to rotate about
its magnetic anisotropy axis. Such a system at rest is described
by the Hamiltonian $\hat{\mathcal{H}}_{\sigma} =
-({\Delta}/{2})\sigma_x$, where $\sigma_x$ is a Pauli matrix and
$\Delta$ is the tunnel splitting of spin-up and spin-down states,
with $z$ being the anisotropy axis. It arises from the terms in
the full Hamiltonian that do not commute with the $z$-component of
the total spin ${\bf S}$. Rotation of the particle by the angle
$\varphi$ about the $Z$-axis transforms the two-state Hamiltonian
into \cite{CG-2010}
\begin{equation}\label{projection'}
\hat{\mathcal{H}}_{\sigma}'  =
-\frac{\Delta}{2}\left[\cos(2S\varphi)\sigma_x +
\sin(2S\varphi)\sigma_y\right]   \,,
\end{equation}
($S$ is dimensionless). Exact eigenvalues of the full Hamiltonian
of a nanomagnet rotating about its anisotropy axis, $\hat{\cal{H}}
= {(\hbar L_z)^2}/{(2I)} + \hat{\mathcal{H}}_{\sigma}'$ (where
$L_z$ is the $z$-component of the dimensionless mechanical angular
momentum and $I$ is the moment of inertia), were obtained in Ref.
\onlinecite{CG-2010}, where it was shown that parameter
\begin{equation}\label{alpha}
\alpha = {2(\hbar S)^2}/(I\Delta)\,,
\end{equation}
determines low energy states of the particle. At $\alpha <
\alpha_1 \equiv \left[1 - 1/(2S)^2\right]^{-1}$ the ground state
and the first excited state are respectively symmetric and
antisymmetric superpositions of ${\bf J} = {\bf L} + {\bf S} = 0$
states shown in Fig. \ref{fig:J0intro}, with energies $E_{\pm} =
\frac{\hbar^2S^2}{2I} \pm \frac{\Delta}{2}$.

Derivation of these results was based upon the assumption that the
parameter $\Delta$ is the same for a stationary magnetic particle
and for a particle that is free to rotate. Instanton method allows
one to test this assumption. Below we find the exact instanton
solution of the equations of motion describing the dynamics of the
magnetic moment and the rotation of the particle. It shows that
mechanical freedom does renormalize the tunnel splitting $\Delta$.
However, this renormalization is small unless $\Delta$ is very
large and $\alpha$ is close to $\alpha_1$.

Consider a high-spin magnetic particle with biaxial anisotropy
that is free to rotate about its easy axis.  The initial state of
the particle is such that its total angular momentum is zero,
$\mathbf{J = S + L} = 0$.  In other words, the total spin
(magnetization) vector points along the easy axis and the particle
rotates about this axis such that these angular momenta are equal
in magnitude and opposite in direction, see Fig.\
\ref{fig:J0intro}. The exchange interaction between individual
spins is strong, so the magnitude of the total spin of the
particle is a constant. The magnetic energy will be expressed
below in terms of $\mathbf M$ which is proportional to the spin,
$\hbar \mathbf S = {\mathbf M}/{\gamma}$. Here $\gamma = - e / 2 m
c <0$ is the electron gyromagnetic ratio. The orbital angular
momentum is associated with the rotational motion of the particle
itself, $\hbar \mathbf L = I \boldsymbol{\dot \varphi}$, where $I$
is the particle's moment of inertia and $\boldsymbol{\dot
\varphi}$ is its angular velocity.

\begin{figure}
\begin{center}
\includegraphics[width=64mm,angle=-90]{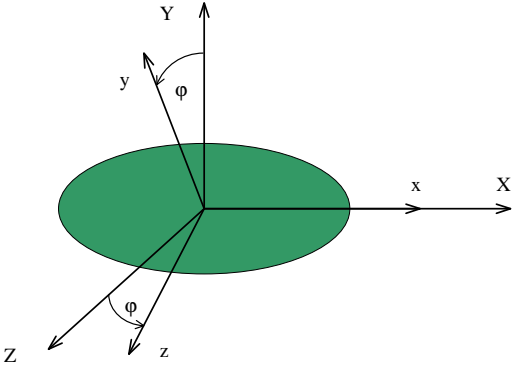}
\end{center}
\caption{ Geometry of particle and laboratory frames. }
\label{fig:rotating}
\end{figure}

We define coordinate systems of the lab frame $(x, y, z)$ and
particle frame $(X, Y, Z)$ as shown in Fig.~\ref{fig:rotating}. In
the particle frame the $x$-axis is along the easy axis in the $xy$
easy plane, and the $z$-axis is the hard axis. The lab frame is
centered at the same origin such that the $X$-axis of the lab
frame coincides with the $x$-axis of the particle frame. The
particle is free to rotate about this axis. At some initial time
$t = 0$ we choose the two coordinate frames to coincide.  The
angle of rotation of the $x$- and $z$-axes with respect to the
$X$- and $Z$-axes is $\varphi (t)$. Notice the change of the easy
axis as compared to the choice of Ref. \onlinecite{CG-2010}, which
is dictated by the mathematics of the problem.

The anisotropy energy is naturally defined in the particle frame:
$\mathcal E_A = k_\perp M_z^2 - k_\parallel M_x^2$. It can be
written in terms of spherical polar coordinates $(\theta, \phi)$
which are defined with respect to the particle-frame axes,
\begin{equation}
\mathcal E_A (\theta, \phi) = \frac{1}{2}\mu_0M_0^2 V [ ( K_\perp
+ K_\parallel \cos^2 \phi ) \cos^2 \theta + K_\parallel \sin^2
\phi]\,. \label{eq:E_A_lattice}
\end{equation}
Here $\mu_0$ is the magnetic permeability of vacuum, $V$ is the
volume of the particle and $M_0 = M / V$ represents the
magnetization which is a constant of the ferromagnetic material.
Anisotropy constants $K_\perp > 0$ and $K_\parallel > 0$ have been
redefined to show the explicit proportionality of the anisotropy
energy to the volume. They are dimensionless numbers, typically of
order unity.

The rotational kinetic energy of the particle is $\mathcal E_{R} =
\frac{1}{2} I \dot {\bm \varphi}^2$. The Lagrangian of our system
consists of the trivial kinetic Lagrangian for the variable ${\bm
\varphi}$,
\begin{equation}\label{Lag-mech}
{\mathcal L}_{L} = \hbar{\bf L}\cdot\dot{\bm \varphi} - \mathcal
E_{R} = \frac{1}{2} I \dot {\bm \varphi}^2\,,
\end{equation}
and the magnetic Lagrangian,
\begin{equation}\label{Lag-S}
{\mathcal L}_{S} = \left ( \frac{M_0 V}{\gamma} \right ) \dot \phi
\cos \theta - \mathcal E'_A (\theta, \phi)\,,
\end{equation}
for the variables $\theta$ and $\phi$. The first term in Eq.\
(\ref{Lag-S}) follows from the fact that $\hbar S_z = \hbar
S\cos\theta$ is the generalized momentum for the coordinate
$\phi$. The second term is the effective magnetic energy in the
rotating frame,
\begin{equation}\label{A-prime}
\mathcal E'_A = \mathcal E_A(\theta, \phi) - \hbar {\bf
S}\cdot\dot{\bm \varphi}\,.
\end{equation}
The last term in this equation is related to the fact that in the
particle frame the rotation is equivalent to the magnetic field
${\bf B} = \dot{\bm \varphi}/\gamma$.

The total Lagrangian of the particle is a sum of ${\mathcal
L}_{L}$ and ${\mathcal L}_{S}$:
\begin{equation}\label{total-L}
\mathcal L = \hbar (\mathbf L + \mathbf S) \cdot \boldsymbol{ \dot
\varphi }
    + \left ( \frac{M_0 V}{\gamma} \right ) \dot \phi \cos \theta
     - \left [ \frac{1}{2} I \boldsymbol{ \dot \varphi }^2 +
    \mathcal E_A (\phi, \theta) \right ]\,.
\end{equation}
The first term reflects the fact that in the presence of a spin
the generator of rotations is ${\bf J} = {\bf L} + {\bf S}$. The
explicit form of the total Lagrangian in terms of the generalized
coordinates $\theta, \phi$, and $\varphi$ is
\begin{eqnarray}
\mathcal L &=& \frac{1}{2} I \boldsymbol{ \dot \varphi }^2
        + \left ( \frac{M_0 V}{\gamma} \right )  \dot \phi \cos \theta
        + \left ( \frac{M_0 V}{\gamma} \right )  \dot \varphi
        \sin \theta  \cos \phi
         \nonumber  \\
    &-& \frac{1}{2}\mu_0 M_0^2 V [ ( K_\perp + K_\parallel \cos^2 \phi )
    \cos^2 \theta + K_\parallel \sin^2 \phi]\,. \nonumber \\
\end{eqnarray}
The equations of motion are Euler-Lagrange equations for $\theta,
\phi$, and $\varphi$:
\begin{eqnarray}
&& \frac{d \phi}{d \tilde t} = (K_\perp + K_\parallel  \cos^2 \phi
) \cos \theta
    +  \dot{\tilde \varphi} \; \frac{\cos \theta  \cos \phi}{\sin \theta}
    \label{eq:phi}\\
&& \frac{d (\cos \theta)}{d \tilde t} = - K_\parallel  \cos \phi
\sin \phi  \sin^2 \theta
     -  \dot{\tilde \varphi}  \sin \theta \sin \phi \label{eq:theta}\\
&& \frac{d}{d \tilde t}  \left[ I \dot{\tilde \varphi} +
\frac{V}{\mu_0 \gamma^2}  \cos \phi  \sin \theta \right ] = 0\,,
\label{eq:EOM_3}
\end{eqnarray}
where we have introduced dimensionless time $\tilde t = \gamma
\mu_0 M_0 t$ and $\dot{\tilde \varphi} = d \varphi / d \tilde t$.

Note that the equations of motion for $\phi$ and $\theta$ can also
be obtained from the Landau-Lifshitz equation with $\mathcal E =
\mathcal E_A'$. Indeed, the equations for $\phi$ and $\theta$ that
follow from Eq.\ (\ref{eq:LL}) (see, e.g., Ref.\
\onlinecite{Lectures}),
\begin{equation}\label{eq:LL_angles}
\frac{\partial \phi}{\partial t} = - \frac{\gamma}{M \sin \theta}
\left ( \frac{\partial \mathcal E_A'}{\partial \theta} \right )\,,
\qquad \frac{\partial \theta}{\partial t} = \frac{\gamma}{M \sin
\theta} \left ( \frac{\partial \mathcal E_A'}{\partial \phi}
\right )\,,
\end{equation}
are identical to the equations (\ref{eq:phi}) and
(\ref{eq:theta}). The third equation of motion, Eq.\
\ref{eq:EOM_3}, is the conservation of the total angular momentum:
\begin{equation}
\frac{d}{d t} [L_X + S_X] = \frac{d}{d t} J_X = 0\,.
\end{equation}
At $J_X = 0$ it is equivalent to the constraint:
\begin{equation}
I \dot \varphi = - \frac{M_0 V}{\gamma} \sin \theta \cos \phi\,.
\label{eq:constraint_0}
\end{equation}
With account of this constraint the equations of motion for $\phi$
and $\theta$ become
\begin{eqnarray}
&& \frac{d \phi}{d \tilde t} = (K_\perp + K_\parallel'  \cos^2
\phi ) \cos \theta \label{phi}\\
&& \frac{d (\cos \theta)}{d \tilde t} = - K_\parallel'  \cos \phi
\sin \phi  \sin^2 \theta\,,\label{theta}
\end{eqnarray}
where
\begin{equation}
K_\parallel' = K_\parallel - K_R, \qquad K_R = \frac{V}{\mu_0
\gamma^2 I} \,. \label{Renorm}
\end{equation}

We see that for ${\bf J} = 0$ the effect of rotations reduces to
the renormalization of the easy axis anisotropy $K_\parallel$.
This is easy to understand from the following consideration. In a
state with $J_x = 0$, equilibrium vectors ${\bf S}$ and ${\bf L}$
look in the opposite directions along the $x$-axis. If ${\bf S}$
deviates from the $x$-axis, $S_x$ decreases and so should $L_x$ to
preserve the condition $J_x = 0$. The decrease of $L_x$
corresponds to the decrease of the rotational energy, $(\hbar
L_x)^2/(2I)$, mandated by $J_x = 0$. Thus, effectively, the
magnetic anisotropy energy associated with the deviation of ${\bf
S}$ from the easy axis becomes smaller when mechanical rotation is
allowed.

We should now look for solutions of equations (\ref{phi}) and
(\ref{theta}). We first notice that
\begin{equation}
\mathcal E = \frac{1}{2}\mu_0M_0^2 V [ ( K_\perp + K_\parallel'
\cos^2 \phi ) \cos^2 \theta
        + K_\parallel' \sin^2 \phi]
\label{eq:E_lab_sub}
\end{equation}
is the integral of motion. This is easy to see by differentiating
this equation on time and substituting in the resulting equation
the time derivatives $\phi$ and $\theta$ from equations
(\ref{phi}) and (\ref{theta}). Not surprisingly, up to a constant,
Eq.\ (\ref{eq:E_lab_sub}) equals the total energy of the particle
in the laboratory frame, $\mathcal E = \mathcal E_A + \frac{1}{2}
I \dot {\bm \varphi}^2$, with account of the constraint
(\ref{eq:constraint_0}). Eq.\ (\ref{phi}) gives
\begin{equation}\label{cos}
\cos \theta = \frac{{d \phi}/{d \tilde t}}{K_\perp + K_\parallel'
\cos^2 \phi}\,.
\end{equation}
This allows one to express $\mathcal E$ in terms of the angle
$\phi$ and its time derivative:
\begin{equation}\label{energy}
\mathcal E = \frac{1}{2}M_0^2 V \left [ \frac{(d \phi / d \tilde
t)^2} {K_\perp + K_\parallel' \cos^2 \phi} + K_\parallel' \sin^2
\phi \right]\,.
\end{equation}
Since this expression is positively defined, the classical energy
minima occur at $\mathcal E = 0$. They correspond to the
stationary magnetization pointing in either direction along the
easy axis, i.e., $\phi = 0, \pi$ with $\cos \theta = 0$ in
accordance with Eq.\ (\ref{cos}).

\begin{figure}
\vspace{24pt}
\includegraphics[width=80mm]{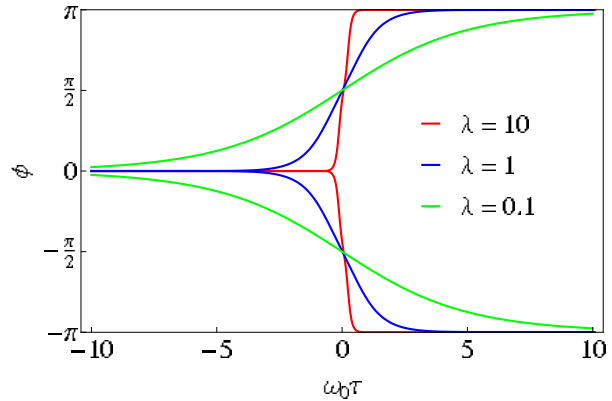}
\caption{Dependence of $\phi$ on $\tau$ for clockwise and
counterclockwise underbarrier rotations of ${\bf M}$.}
\label{fig:instantons}
\end{figure}

Equation $\mathcal E = 0$ has no real-time solutions for $\phi$
that connect the two degenerate classical energy minima. However,
in imaginary time, $\tilde \tau = i \tilde t$, equation $\mathcal
E = 0$ is equivalent to
\begin{equation}
\left ( \frac{d \phi}{d \tilde \tau} \right )^2 = K_\parallel'
\sin^2 \phi \; (K_\perp + K_\parallel'  \cos^2 \phi)\,.
\end{equation}
Such equation has instanton solutions that connect the classical
energy minima:
\begin{equation}
\phi (\tau) = \pm \text{arccos} \left \{
    - \frac{ \sinh (\omega_0 \tau) }{ \sqrt{ \lambda + \cosh^2 (\omega_0 \tau)} }
    \right \}\,.
\label{eq:phi_tau}
\end{equation}
The $\tau$-dependence of $\theta$ is given by Eq.\ (\ref{cos}),
and the $\tau$-dependence of $\varphi$ is given by Eq.\
(\ref{eq:constraint_0}). Here
\begin{equation}
\lambda = {K_\parallel'}/{K_\perp}, \qquad
    \omega_0 =  |\gamma| \mu_0 M_0 \sqrt{ K_\parallel' ( K_\parallel' + K_\perp ) }\,.
\end{equation}
The positive and negative signs correspond to the two possible
trajectories, which are counterclockwise and clockwise rotations
of the magnetization from $\phi = 0$ at $\tau = - \infty$ to $\phi
= \pm \pi$ at $\tau = + \infty$, respectively, see Fig.\
\ref{fig:instantons}.

The tunnel splitting has the form $\Delta = A e^B$, where A is of
the order of quantized oscillations near the minimum of the
potential well and
\begin{equation}
B
    = \frac{i}{\hbar}  \int_{- \infty}^{+\infty} dt \mathcal L
\end{equation}
is the WKB exponent. Substituting here $\mathcal L$ of Eq.\
(\ref{total-L}) at ${\bf J} = 0$, one obtains for the instanton
trajectory
\begin{equation}
B = - S \ln \left( \frac{ \sqrt{ 1 + \lambda } +
    \sqrt \lambda }{ \sqrt{ 1 + \lambda } - \sqrt \lambda }
    \right )\,.\label{WKB}
\end{equation}

To see the effect of the mechanical freedom of the particle on
spin tunneling we define a dimensionless parameter $\alpha' =
{K_R}/{K_\parallel}$. In a microscopic theory the easy-axis
crystal field is presented as $-DS_z^2$. The connection between
$D$ and the parameter $K_{\parallel}$ of the macroscopic theory is
\cite{Lectures}
\begin{equation}
K_{\parallel} = \left(2-\frac{1}{s}\right) \frac{D V_0}{\mu_0
(\hbar \gamma)^2}\,,
\end{equation}
where $s$ is spin per unit cell of the crystal and $V_0$ is the
volume of the unit cell. (Singularity at $s = 1/2$ reflects the
fact that single-ion magnetic anisotropy does not exist for spin
$1/2$). The total spin of a ferromagnetic particle can be
presented as $S = s(V /V_0)$. Consequently,
\begin{equation}
\alpha' = \frac{S \hbar^2}{(2s-1)I D} =
\frac{\Delta}{(2s-1)2SD}\,\alpha\,,
\end{equation}
where we have used Eqs.\ (\ref{alpha}) and (\ref{Renorm}).
Renormalization of the easy-axis anisotropy by rotations can be
presented in the form $K_\parallel' = K_\parallel(1-\epsilon)$,
where
\begin{equation}
\epsilon =
\left(\frac{\alpha}{2s}\right)\left(\frac{1-\frac{1}{2S}}{1-\frac{1}{2s}}\right)
\frac{\Delta}{E_1} \label{epsilon}
\end{equation}
and $E_1 = (2S-1)D$ is the energy of the first excited spin state
at $\Delta = 0$. The low energy limit that we are studying
corresponds to $\Delta \ll E_1$ and $\alpha < \alpha_1 \equiv
\left[1 - 1/(2S)^2\right]^{-1}$, see Ref. \onlinecite{CG-2010}. In
this limit $\epsilon$ is small. Consider, e.g., the case of large
$S$ and large $\lambda$ (small tunneling rate). According to Eq.\
(\ref{WKB}) in this case $B = -S\ln(4\lambda)$ so that $\Delta
\propto \exp{[-S \ln (4\lambda)]}$. It is easy to see from this
expression that mechanical rotation renormalizes $\Delta$ by a
factor $\exp{(\epsilon S)}$. Normally it would not be large
compared to one. However, since small $\epsilon$ in the exponent
is multiplied by a large $S$, it is not out of question that at
sufficiently large $\Delta$ a slight increase of the tunnel
splitting would be observable in spin clusters that are free to
rotate.

This work has been supported by the NSF grant No. DMR-0703639.


\begin{thebibliography}{99}

\bibitem{LL-old}
L. D. Landau and E. M. Lifshitz, Phys. Zs. Sowjet. {\bf 8}, 153
(1935)

\bibitem{Lectures}
E. M. Chudnovsky and J. Tejada, {\it Lectures on Magnetism}
(Rinton Press, Princeton, NJ, 2008).

\bibitem{EC-JETP}
E. M. Chudnovsky, Sov. Phys. JETP {\bf 50}, 1035 (1979).

\bibitem{EC-Gunther}
E. M. Chudnovsky and L. Gunther, Phys. Rev. Lett. {\bf 60}, 661
(1988).

\bibitem{MQT}
E. M. Chudnovsky and J. Tejada, {\it Macroscopic Quantum Tunneling
of the Magnetic Moment } (Cambridge University Press, Cambridge,
UK, 1998).

\bibitem{Tejada-2010}
J. Tejada, R. D. Zysler, E. Molins, and E. M. Chudnovsky, Phys.
Rev. Lett. {\bf 104}, 027202 (2010).

\bibitem{Cox}
D. M. Cox, D. J. Trevor, R. L. Whetten, E. A. Rohlfing, and A.
Kaldor, Phys. Rev. B {\bf 32}, 7290 (1985).

\bibitem{Heer}
W. A. de Heer, P. Milani, and A. Chatelain, Phys. Rev. Lett. {\bf
65}, 488 (1990).

\bibitem{Bucher}
J. P. Bucher, D. C. Douglass, L. A. Bloomfield, Phys. Rev. Lett.
{\bf 66}, 3052 (1991).

\bibitem{Douglass}
D. C. Douglass, D. M. Cox, J. P. Buchwer, L. A. Bloomfield, Phys.
Rev. B {\bf 47}, 12874 (1993).

\bibitem{Billas}
I. M. L. Billas, J. A. Becker, A. Chatalain, W. A. de Heer, Phys.
Rev. Lett. {\bf 71}, 4067 (1993).

\bibitem{Xu}
X. Xu, S. Yin, R. Moro, and W. A. de Heer, Phys. Rev. Lett. {\bf
95}, 237209 (2005).

\bibitem{Zobbi}
L. Zobbi, M. Mannini, M. Pacchioni, G. Chastanet, D. Bonacchi, C.
Zanardi, R. Biagi, U. del Pennino, D. Gatteschi, A. Cornia, and R.
Sessoli, Chem. Comm. {\bf 12}, 1640 (2005).

\bibitem{Martinez}
R. V. Mart\'{i}nez, F. Garc\'{i}a, R. Garc\'{i}a, E. Coronado, A.
Forment-Aliaga, F. M. Romero, and S. Tatay, Adv. Mater. {\bf 19},
291 (2007).

\bibitem{Barraza}
S. Barraza-Lopez, M. C. Avery, and K. Park, Phys. Rev. B {\bf 76},
224413 (2007).

\bibitem{Pennino}
U. del Pennino, V. Corradini, R. Biagi, V. De Renzi, F. Moro, D.
W. Boukhvalov, G. Panaccione, M. Hochstrasser, C. Carbone, C. J.
Milios, and E. K. Brechin, Phys. Rev. B {\bf 77}, 085419 (2008).

\bibitem{Wern-CNT}
J.-P. Cleuziou, W. Wernsdorfer, V. Bouchiat, T. Ondarcuhu, and M.
Monthioux, Nature Nanotechnology {\bf 1}, 53 (2006); J.-P.
Cleuziou, W. Wernsdorfer, S. Andergassen, S. Florens, V. Bouchiat,
T. Ondarcuhu, and M. Monthioux, Phys. Rev. Lett. {\bf 99}, 117001
(2007); L. Bogani, R. Maurand, L. Marty, C. Sangregorio, C.
Altavillad, and W. Wernsdorfer, J. Mater. Chem. {\bf 20}, 2099
(2010).

\bibitem{Heersche}
H. B. Heersche, Z. de Groot, J. A. Folk, H. S. van der Zant, C.
Romeike, M. R. Wegewijs, L. Zobbi, D. Barreca, E. Tondello, and A.
Cornia, Phys. Rev. Lett. {\bf 96}, 206801 (2006).

\bibitem{Jo}
M.-H. Jo, J. E. Grose, K. Baheti, M. M. Deshmukh, J. J. Sokol, E.
M. Rumberger, D. N. Hendrickson, J. R. Long, H. Park, and D. C.
Ralph, Nano Lett. {\bf 6}, 2014 (2006).

\bibitem{Henderson}
J. J. Henderson, C. M. Ramsey, E. del Barco, A. Mishra, and G.
Christou, J. Appl. Phys. {\bf 101}, 09E102 (2007).

\bibitem{Voss}
S. Voss, M. Fonin, U. Rudiger, M. Burgert, and U. Groth, Phys.
Rev. B {\bf 78}, 155403 (2008).

\bibitem{Park}
S. Barraza-Lopez, K. Park, V. Garc\'{i}a-Su\'{a}rez,and J. Ferrer,
Phys. Rev. Lett. {\bf 102}, 246801 (2009).

\bibitem{CG-2010}
E. M. Chudnovsky and D. A. Garanin, Pys. Rev. B {\bf 81}, 214423
(2010)

\bibitem{Jaafar}
R. Jaafar, E. M. Chudnovsky, and D. A. Garanin, Europhys. Lett.
{\bf 89}, 27001 (2010).

\bibitem{LCT}
S. Lend\'{i}nez, E. M. Chudnovsky, and J. Tejada, cond-mat, arXiv:
1008.2142, to appear in Physical Review B.

\bibitem{Kovalev}
A. A. Kovalev, L. X. Hayden, G. E. W. Bauer, and Y. Tserkovnyak,
cond-mat, arXiv: 1011.2242.

\bibitem{Edmonds}
A. R. Edmonds, {\it Angular Momentum in Quantum Mechanics}
(Princeton University Press, Princeton, New Jersey, 1957).

\end{thebibliography}
\end{document}